\documentclass[letterpaper,aps,prl,superscriptaddress,showpacs,floatfix,twocolumn]{revtex4}

\usepackage{graphicx}	
\usepackage{verbatim}   

\usepackage{xspace}	

\newcommand{\pt}{\mbox{$p_T$}\xspace}
\newcommand{\ptt}{\mbox{$p_T^t$}\xspace}
\newcommand{\pta}{\mbox{$p_T^a$}\xspace}
\newcommand{\p}{\mbox{$\pi^{0}$}\xspace}
\newcommand{\ps}{\mbox{$\pi^{0}{\rm s}$}\xspace}
\newcommand{\h}{\mbox{$h^{\pm}$}\xspace}
\newcommand{\raa}{\mbox{$R_{\rm AA}$}\xspace}
\newcommand{\iaa}{\mbox{$I_{\rm AA}$}\xspace}

\newcommand \sqsn{\mbox{$\sqrt{s_{_{NN}}}$}\xspace}

\begin{document}



\title{ Trends in Yield and Azimuthal Shape Modification \\
in Dihadron Correlations in Relativistic Heavy Ion Collisions
}

\newcommand{\abilene}{Abilene Christian University, Abilene, Texas 79699, USA}
\newcommand{\acadsin}{Institute of Physics, Academia Sinica, Taipei 11529, Taiwan}
\newcommand{\banaras}{Department of Physics, Banaras Hindu University, Varanasi 221005, India}
\newcommand{\barc}{Bhabha Atomic Research Centre, Bombay 400 085, India}
\newcommand{\bnlcoll}{Collider-Accelerator Department, Brookhaven National Laboratory, Upton, New York 11973-5000, USA}
\newcommand{\bnlphys}{Physics Department, Brookhaven National Laboratory, Upton, New York 11973-5000, USA}
\newcommand{\caucr}{University of California - Riverside, Riverside, California 92521, USA}
\newcommand{\charlesczech}{Charles University, Ovocn\'{y} trh 5, Praha 1, 116 36, Prague, Czech Republic}
\newcommand{\chonbuk}{Chonbuk National University, Jeonju, 561-756, Korea}
\newcommand{\ciae}{China Institute of Atomic Energy (CIAE), Beijing, People's Republic of China}
\newcommand{\cns}{Center for Nuclear Study, Graduate School of Science, University of Tokyo, 7-3-1 Hongo, Bunkyo, Tokyo 113-0033, Japan}
\newcommand{\colorado}{University of Colorado, Boulder, Colorado 80309, USA}
\newcommand{\columbia}{Columbia University, New York, New York 10027 and Nevis Laboratories, Irvington, NY 10533, USA}
\newcommand{\czechtech}{Czech Technical University, Zikova 4, 166 36 Prague 6, Czech Republic}
\newcommand{\dapnia}{Dapnia, CEA Saclay, F-91191, Gif-sur-Yvette, France}
\newcommand{\debrecen}{Debrecen University, H-4010 Debrecen, Egyetem t{\'e}r 1, Hungary}
\newcommand{\elte}{ELTE, E{\"o}tv{\"o}s Lor{\'a}nd University, H - 1117 Budapest, P{\'a}zm{\'a}ny P. s. 1/A, Hungary}
\newcommand{\ewha}{Ewha Womans University, Seoul 120-750, Korea}
\newcommand{\fit}{Florida Institute of Technology, Melbourne, Florida 32901, USA}
\newcommand{\fsu}{Florida State University, Tallahassee, Florida 32306, USA}
\newcommand{\gsu}{Georgia State University, Atlanta, Georgia 30303, USA}
\newcommand{\hiroshima}{Hiroshima University, Kagamiyama, Higashi-Hiroshima 739-8526, Japan}
\newcommand{\ihepprot}{IHEP Protvino, State Research Center of Russian Federation, Institute for High Energy Physics, Protvino, 142281, Russia}
\newcommand{\illuiuc}{University of Illinois at Urbana-Champaign, Urbana, Illinois 61801, USA}
\newcommand{\instpasczech}{Institute of Physics, Academy of Sciences of the Czech Republic, Na Slovance 2, 182 21 Prague 8, Czech Republic}
\newcommand{\isu}{Iowa State University, Ames, Iowa 50011, USA}
\newcommand{\jinrdubna}{Joint Institute for Nuclear Research, 141980 Dubna, Moscow Region, Russia}
\newcommand{\jyvaskyla}{Helsinki Institute of Physics and University of Jyv{\"a}skyl{\"a}, P.O.Box 35, FI-40014 Jyv{\"a}skyl{\"a}, Finland}
\newcommand{\kek}{KEK, High Energy Accelerator Research Organization, Tsukuba, Ibaraki 305-0801, Japan}
\newcommand{\kfki}{KFKI Research Institute for Particle and Nuclear Physics of the Hungarian Academy of Sciences (MTA KFKI RMKI), H-1525 Budapest 114, POBox 49, Budapest, Hungary}
\newcommand{\korea}{Korea University, Seoul, 136-701, Korea}
\newcommand{\kurchatov}{Russian Research Center ``Kurchatov Institute", Moscow, Russia}
\newcommand{\kyoto}{Kyoto University, Kyoto 606-8502, Japan}
\newcommand{\labllr}{Laboratoire Leprince-Ringuet, Ecole Polytechnique, CNRS-IN2P3, Route de Saclay, F-91128, Palaiseau, France}
\newcommand{\lawllnl}{Lawrence Livermore National Laboratory, Livermore, California 94550, USA}
\newcommand{\losalamos}{Los Alamos National Laboratory, Los Alamos, New Mexico 87545, USA}
\newcommand{\lpc}{LPC, Universit{\'e} Blaise Pascal, CNRS-IN2P3, Clermont-Fd, 63177 Aubiere Cedex, France}
\newcommand{\lund}{Department of Physics, Lund University, Box 118, SE-221 00 Lund, Sweden}
\newcommand{\maryland}{University of Maryland, College Park, Maryland 20742, USA}
\newcommand{\mass}{Department of Physics, University of Massachusetts, Amherst, MA 01003-9337, USA }
\newcommand{\muenster}{Institut fur Kernphysik, University of Muenster, D-48149 Muenster, Germany}
\newcommand{\muhlenberg}{Muhlenberg College, Allentown, Pennsylvania 18104-5586, USA}
\newcommand{\myongji}{Myongji University, Yongin, Kyonggido 449-728, Korea}
\newcommand{\nagasaki}{Nagasaki Institute of Applied Science, Nagasaki-shi, Nagasaki 851-0193, Japan}
\newcommand{\newmex}{University of New Mexico, Albuquerque, New Mexico 87131, USA }
\newcommand{\nmsu}{New Mexico State University, Las Cruces, New Mexico 88003, USA}
\newcommand{\ornl}{Oak Ridge National Laboratory, Oak Ridge, Tennessee 37831, USA}
\newcommand{\orsay}{IPN-Orsay, Universite Paris Sud, CNRS-IN2P3, BP1, F-91406, Orsay, France}
\newcommand{\peking}{Peking University, Beijing, People's Republic of China}
\newcommand{\pnpi}{PNPI, Petersburg Nuclear Physics Institute, Gatchina, Leningrad region, 188300, Russia}
\newcommand{\riken}{RIKEN Nishina Center for Accelerator-Based Science, Wako, Saitama 351-0198, JAPAN}
\newcommand{\rikjrbrc}{RIKEN BNL Research Center, Brookhaven National Laboratory, Upton, New York 11973-5000, USA}
\newcommand{\rikkyo}{Physics Department, Rikkyo University, 3-34-1 Nishi-Ikebukuro, Toshima, Tokyo 171-8501, Japan}
\newcommand{\saispbstu}{Saint Petersburg State Polytechnic University, St. Petersburg, Russia}
\newcommand{\saopaulo}{Universidade de S{\~a}o Paulo, Instituto de F\'{\i}sica, Caixa Postal 66318, S{\~a}o Paulo CEP05315-970, Brazil}
\newcommand{\seoulnat}{System Electronics Laboratory, Seoul National University, Seoul, Korea}
\newcommand{\stonybrkc}{Chemistry Department, Stony Brook University, Stony Brook, SUNY, New York 11794-3400, USA}
\newcommand{\stonycrkp}{Department of Physics and Astronomy, Stony Brook University, SUNY, Stony Brook, New York 11794, USA}
\newcommand{\subatech}{SUBATECH (Ecole des Mines de Nantes, CNRS-IN2P3, Universit{\'e} de Nantes) BP 20722 - 44307, Nantes, France}
\newcommand{\tenn}{University of Tennessee, Knoxville, Tennessee 37996, USA}
\newcommand{\titech}{Department of Physics, Tokyo Institute of Technology, Oh-okayama, Meguro, Tokyo 152-8551, Japan}
\newcommand{\tsukuba}{Institute of Physics, University of Tsukuba, Tsukuba, Ibaraki 305, Japan}
\newcommand{\vandy}{Vanderbilt University, Nashville, Tennessee 37235, USA}
\newcommand{\waseda}{Waseda University, Advanced Research Institute for Science and Engineering, 17 Kikui-cho, Shinjuku-ku, Tokyo 162-0044, Japan}
\newcommand{\weizmann}{Weizmann Institute, Rehovot 76100, Israel}
\newcommand{\yonsei}{Yonsei University, IPAP, Seoul 120-749, Korea}
\affiliation{\abilene}
\affiliation{\acadsin}
\affiliation{\banaras}
\affiliation{\barc}
\affiliation{\bnlcoll}
\affiliation{\bnlphys}
\affiliation{\caucr}
\affiliation{\charlesczech}
\affiliation{\chonbuk}
\affiliation{\ciae}
\affiliation{\cns}
\affiliation{\colorado}
\affiliation{\columbia}
\affiliation{\czechtech}
\affiliation{\dapnia}
\affiliation{\debrecen}
\affiliation{\elte}
\affiliation{\ewha}
\affiliation{\fit}
\affiliation{\fsu}
\affiliation{\gsu}
\affiliation{\hiroshima}
\affiliation{\ihepprot}
\affiliation{\illuiuc}
\affiliation{\instpasczech}
\affiliation{\isu}
\affiliation{\jinrdubna}
\affiliation{\jyvaskyla}
\affiliation{\kek}
\affiliation{\kfki}
\affiliation{\korea}
\affiliation{\kurchatov}
\affiliation{\kyoto}
\affiliation{\labllr}
\affiliation{\lawllnl}
\affiliation{\losalamos}
\affiliation{\lpc}
\affiliation{\lund}
\affiliation{\maryland}
\affiliation{\mass}
\affiliation{\muenster}
\affiliation{\muhlenberg}
\affiliation{\myongji}
\affiliation{\nagasaki}
\affiliation{\newmex}
\affiliation{\nmsu}
\affiliation{\ornl}
\affiliation{\orsay}
\affiliation{\peking}
\affiliation{\pnpi}
\affiliation{\riken}
\affiliation{\rikjrbrc}
\affiliation{\rikkyo}
\affiliation{\saispbstu}
\affiliation{\saopaulo}
\affiliation{\seoulnat}
\affiliation{\stonybrkc}
\affiliation{\stonycrkp}
\affiliation{\subatech}
\affiliation{\tenn}
\affiliation{\titech}
\affiliation{\tsukuba}
\affiliation{\vandy}
\affiliation{\waseda}
\affiliation{\weizmann}
\affiliation{\yonsei}
\author{A.~Adare} \affiliation{\colorado}
\author{S.~Afanasiev} \affiliation{\jinrdubna}
\author{C.~Aidala} \affiliation{\mass}
\author{N.N.~Ajitanand} \affiliation{\stonybrkc}
\author{Y.~Akiba} \affiliation{\riken} \affiliation{\rikjrbrc}
\author{H.~Al-Bataineh} \affiliation{\nmsu}
\author{J.~Alexander} \affiliation{\stonybrkc}
\author{T.~Alho} \affiliation{\jyvaskyla}
\author{K.~Aoki} \affiliation{\kyoto} \affiliation{\riken}
\author{L.~Aphecetche} \affiliation{\subatech}
% ...
\collaboration{PHENIX Collaboration} \noaffiliation

\date{\today}

\begin{abstract}
Fast parton probes produced by hard scattering and embedded within 
collisions of large nuclei have shown that partons suffer large 
energy loss and that the produced medium may respond collectively 
to the lost energy. We present measurements of neutral pion 
trigger particles at transverse momenta \ptt = 4--12 GeV/$c$ and 
associated charged hadrons (\pta = 0.5--7 GeV/$c$) as a function 
of relative azimuthal angle $\Delta\phi$ at midrapidity in Au$\mathord{+}$Au and 
$p\mathord{+}p$ collisions at \sqsn = 200 GeV.  These data lead to 
two major observations.  First, the relative angular distribution 
of low momentum hadrons, whose shape modification has been 
interpreted as a medium response to parton energy loss, is found 
to be modified for \ptt $<$ 7 GeV/$c$. At higher \ptt, the data are
consistent with unmodified or very weakly modified shapes,
even for the lowest measured 
\pta. This observation presents a quantitative challenge to medium 
response scenarios.  Second, the associated yield of hadrons 
opposite to the trigger particle in Au$\mathord{+}$Au relative to 
that in $p\mathord{+}p$ (\iaa) is found to be suppressed at large 
momentum (\iaa $\approx$ 0.35--0.5), but less than the single 
particle nuclear modification factor (\raa $\approx$ 0.2).
\end{abstract}

	


\maketitle


Angular correlations between the hadronic fragments of energetic 
partons are an essential tool for understanding the hot dense 
matter produced in relativistic heavy ion 
collisions~\cite{Adams:2005ph,ppg032,ppg067,starhighpt,ppg083,Abelev:2009qa}. 
It is expected that fast partons dissipate a large portion of 
their energy while traversing this medium, and that correlations 
between the hadronic fragments of these partons reflect the 
influence of the energy loss and its deposition into the medium.  
It has already been observed in 
dihadron correlations from central Au+Au collisions that 
both the shape of the relative azimuthal angular distribution and 
the yield of jet-like fragment pairs can depart significantly from those of 
$p\mathord{+}p$ collisions~\cite{Adams:2005ph,ppg083}. The 
underlying mechanisms for jet modification are not yet fully 
understood, but partonic energy loss by QCD radiative processes 
and collisions with medium constituents, as well as the evolution 
of the lost energy, should contribute to the modification of 
single and pair yields of hadrons associated with jets.

In the moderate \ptt, \pta range ($\sim$ 2--5 GeV/$c$), a 
pronounced away-side peak broadening~\cite{ppg032} and shape 
modification~\cite{ppg067, ppg083} have been observed. The modified 
shape has been interpreted in some models as a medium response to 
the energy deposited by partons. These include large-angle gluon 
radiation~\cite{largAng1, largAng2}, \v{C}erenkov gluon 
radiation~\cite{cherenkov}, and Mach-shock or wave 
excitations~\cite{mach, Gubser:2007ga}.  Alternative explanations 
include fluctuating background 
correlations~\cite{ps,Takahashi:2009na} and jets deflected by the 
medium~\cite{deflect}.

Previous measurements~\cite{starhighpt, ppg083} at \ptt, \pta 
$\gtrsim$ 5 GeV/$c$ have shown that away-side correlations exhibit 
suppressed jet peaks with shapes similar to those observed in $p\mathord{+}p$ collisions.
The resemblance to $p\mathord{+}p$ at the highest 
momenta \ptt and \pta may be indicative of 
selective sensitivity to parton pairs that are emitted tangentially 
near the medium surface and thus suffer minimal energy loss, or
alternatively, that some energetic partons lose significant energy in medium,
but the effect from such cases are only visisble at very low \pta.
However, these high-\pt 
results (\pt $\gtrsim$ 5 GeV/$c$) are averaged over broad momentum 
ranges to cope with statistical limitations.


The results presented here are based on minimum-bias 
Au$\mathord{+}$Au and 
photon-triggered~\cite{PhysRevLett.91.241803} $p\mathord{+}p$ 
collisions at \sqsn=200 GeV collected with the PHENIX detector in 
2006 and 2007. The event centrality in Au$\mathord{+}$Au is 
determined by categorizing the integrated charge seen by the 
beam-beam counters~\cite{bbcref} by upper percentile.  After the 
application of event quality cuts, 3.24 million level-1 ``photon'' 
triggered $p\mathord{+}p$ events and 1.78 billion minimum-bias 
Au$\mathord{+}$Au events were used in this analysis.

Neutral pion triggers are reconstructed from photon clusters 
measured by lead-glass and lead-scintillator electromagnetic 
calorimeters in the two central arms of PHENIX, covering $|\eta| < 
0.35$ and $2\times90^{\circ}$ in azimuth~\cite{ppg080}. Neutral 
pions are identified in each event through 2$\gamma$ decay by 
pairing all photons satisfying a minimum energy threshold cut and 
requiring the reconstructed mass to lie near the \p mass peak. 
More restrictive cuts are used in more central events and for 
lower-\pt \ps to reduce the rate of random associations and 
preserve a \p identification signal-to-background ratio (S/B) 
larger than 4:1 for central Au$\mathord{+}$Au and 20:1 in 
$p\mathord{+}p$. A systematic uncertainty of $<$1--6\%, depending 
on S/B, is included for the \p signal extraction.

Charged hadron partners are reconstructed in the central arms 
using the drift chambers (DC) with hit association requirements in 
two layers of multi-wire proportional chambers with pad readout 
(PC1 and PC3), achieving a momentum resolution of $0.7\% \oplus 
1.1\% p$ (GeV/$c$).  Only tracks with full and unambiguous DC and 
PC1 hit information are used.  Projections of these tracks are 
required to match a PC3 hit within a $\pm 2 \sigma$ proximity 
window to reduce background from conversion and decay products.

All trigger-partner pairs satisfying the identification 
requirements within an event are measured. These pairs are 
corrected for the PHENIX acceptance through a process of event 
mixing, and then background pairs which are correlated through the 
reaction plane are subtracted. The conditional jet pair 
multiplicity per trigger particle is thus determined by:
\begin{eqnarray}
  \frac{1}{N^{t}} \frac{dN^{\rm pair}}{d\Delta\phi} &=&
  \frac{N^{a}}{2\pi\epsilon^{a}}
  \left[\frac{dN^{\rm pair}_{same}/d\Delta\phi}{dN^{\rm pair}_{\rm mix}/d\Delta\phi} \right.
    \nonumber \\
    & & \left. - \xi
    \left( 1 + 2 \langle v^{t}_2 v^{a}_2 \rangle \cos
      \left(2\Delta\phi\right) \right) \rule{0mm}{4.3mm} \right],
\centering
\end{eqnarray}
where $N^{t}$ ($N^{a}$) is the number of trigger (associated) particles~\cite{ppg083}.
The background modulation accounts for quadrupole anisotropy only, 
and is assumed to factorize such that $\langle v^{t}_2 v^{a}_2 
\rangle \approx \langle v^{t}_2 \rangle \langle v^{a}_2 
\rangle$~\cite{ppg067}. The elliptic flow coefficients, $v_2$, are 
taken from recent PHENIX measurements of neutral pions~\cite{wei} 
and charged hadrons~\cite{ppg098}. The background level, $\xi$, is 
determined in Au$\mathord{+}$Au collisions using the absolute 
background subtraction method~\cite{abs}.  A pedestal subtraction 
employing the zero-yield-at-minimum (ZYAM) method is used in 
$p\mathord{+}p$. In certain cases, {\it e.g.} very broad jets, the 
ZYAM method could lead to an over-subtraction by removing signal 
pairs. The effect is typically small in $p\mathord{+}p$ where an 
additional 6\% global scale uncertainty is applied. Charged hadron 
acceptance and efficiency corrections, $\epsilon^{a}$, are 
calculated via full detector simulations~\cite{ppg083}.

\begin{figure}[tb]
  \centering
  \includegraphics[width=1.0\linewidth]{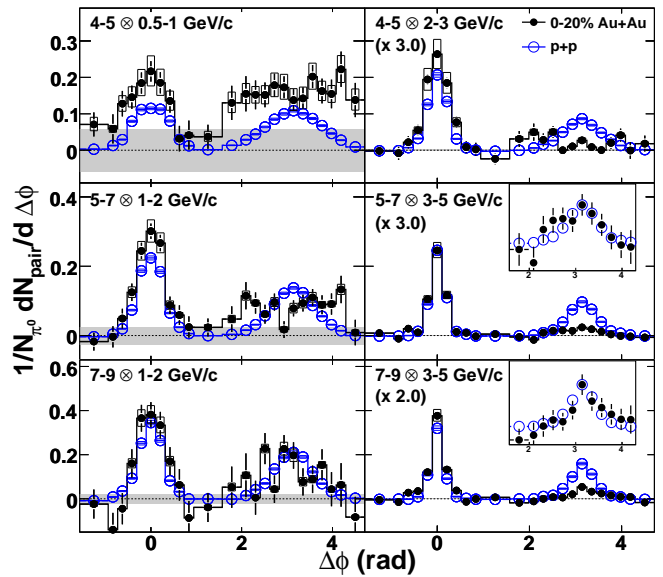}
  \caption{\label{fig:jet}
    (Color online) Per-trigger jet pair yield vs. $\Delta\phi$ for
    selected \p trigger and $h^\pm$ partner $p_T$ combinations
    (\ptt $\otimes$ \pta) in Au$\mathord{+}$Au and $p\mathord{+}p$
    collisions (solid and open symbols, respectively). Depicted
    Au$\mathord{+}$Au systematic uncertainties include point-to-point
    correlated background level and modulation uncertainties (gray
    bands and open boxes, respectively). For shape comparison insets
    show away-side distributions scaled to match at $\Delta\phi =
    \pi$. 
  }
\end{figure}

Figure~\ref{fig:jet} shows 
the resulting per-trigger jet pair yields for selected 
trigger-partner combinations in $p\mathord{+}p$ and the 20\% most 
central Au$\mathord{+}$Au collisions.
On the near side, the widths in central Au$\mathord{+}$Au are 
comparable to $p\mathord{+}p$ over the full \ptt and \pta ranges, 
while the yields are slightly enhanced at low \pt, matching 
$p\mathord{+}p$ as \pt increases.   On the opposing side, qualitatively one observes that for low \ptt
and low \pta the Au$\mathord{+}$Au jet peaks are strongly broadened and 
non-Gaussian.  In contrast, at high \ptt and high \pta the yield is 
substantially suppressed, but the shape appears consistent with the
measurement in the $p\mathord{+}p$ case (as has been previously reported in much
broader \pt ranges for unidentified charged hadron triggers~\cite{ppg083,starhighpt}).
Here we quantify the trends in the shape and yield between these two extremes.

\begin{figure}[tb]
  \centering
  \includegraphics[width=1.0\linewidth]{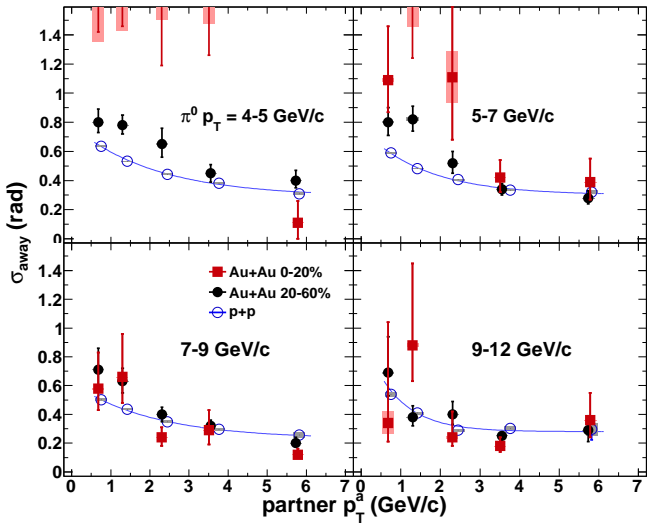}
  \caption{\label{fig:sigma} (Color online) Away-side jet widths from
    a Gaussian fit by $h^\pm$ partner momentum for various \p trigger
    momenta in $p\mathord{+}p$ (open circles), midcentral 20--60\%
    Au$\mathord{+}$Au (solid circles), and central 0--20\%
    Au$\mathord{+}$Au collisions (squares). For comparison, an
    interpolation of the $p\mathord{+}p$ is depicted (curve). In cases
    where the best fit $\sigma_{\rm away} > \pi/2$ radians, the 
    point is off the plot.
  }
\end{figure}

First, we have performed a fit to the away-side distribution over the
range $|\Delta \phi - \pi| < \pi/2$ to a simple Gaussian distribution.
Figure~\ref{fig:sigma} shows the results.
In $p\mathord{+}p$ collisions, the away-side width narrows at 
higher trigger and partner momentum as expected from the angular 
ordering of jet fragmentation. 
For \ptt $>$ 7 GeV/$c$, the widths are consistent
within uncertainties between $p\mathord{+}p$ and Au$\mathord{+}$Au at
all \pta.  
There is no evidence of large jet broadening from in-medium 
scattering~\cite{deflect} or from initial state 
effects~\cite{heavyIonKT}, expected for surviving partons produced 
in the interior rather than the surface of the medium.  However, it is
also possible that for high \ptt the broadening is modest for the 
leading parton and its fragmentation products and the radiated energy
results in only very low \pta hadrons (mostly with \pta $<$ 0.5 GeV/c).


For \ptt $<$ 7 GeV/$c$, the away-side widths are
significantly wider than in $p\mathord{+}p$, except at the highest \pta.
Note that for \ptt $<$ 7 GeV/$c$ and low \pta, the best fit 
$\sigma_{\rm away}$ values are
larger than $\pi/2$ radians.  
These trends in shape are further quantified with the use of a 
$\chi^2$ test to examine the hypothesis that the central 
Au$\mathord{+}$Au jet shape on the near and away side is the same 
as the $p\mathord{+}p$ jet shape.   For \ptt $>$7 GeV/$c$, agreement 
is found for all \pta. However, for \ptt at 5--7 (4--5) GeV/$c$, 
the agreement worsens sharply for \pta $<$ 3 (4) GeV/$c$ 
as the away-side jet becomes increasingly broad.  For example, the p-values for agreement between the
$p\mathord{+}p$ and Au$\mathord{+}$Au shapes for \pta = 1-2 GeV/$c$
are very small ($<$ $10^{-4}$) for \ptt = 4--5 and 5--7 GeV/$c$, but indicate
reasonable agreement (0.33 and 0.16) for \ptt = 7--9 and 9--12 GeV/$c$, respectively.  
The statistical precision of the experimental data does not allow
conclusion of a sharp transition in the shape; 
however, there is a clear indication of
a trend towards either much smaller modification or unmodifed jet shapes
for higher \ptt at all \pta.  To confirm this finding, we compared the away-side distributions
in Au$\mathord{+}$Au central events for \ptt 5--7 GeV/$c$ with \ptt 7--9 GeV/$c$
for \pta 1--2 GeV/$c$ (see Fig. 1) and find the probability
that they have a common source is small (p-value $<$ 0.07).



The lack of large away-side shape modification for \ptt $>$ 7 GeV/$c$ and \pta $<$ 3 GeV/$c$ is surprising as medium 
response effects are not generally expected to decrease at larger 
\ptt.  In descriptions where the medium-induced energy loss 
($\Delta E$) is nearly proportional to the initial parton energy 
($E$)~\cite{Kharzeev:2008he}, and where the lost energy produces a 
medium response, a larger medium modification is expected for 
higher momentum partons.  Within our statistical precision, no 
evidence for this is seen; rather, the opposite is found. However, 
should $\Delta E/E$ fall steeply with increasing parton \pt, an 
increased contribution from partons which have lost little energy 
could make an observation of the medium response more difficult.  
In alternative models of fluctuating background 
correlations~\cite{ps,Takahashi:2009na}, the modification is 
predicted to diminish at higher trigger \pt as the background 
contribution drops, in agreement with observations.

\begin{figure}[tb]
  \centering
  \includegraphics[width=1.0\linewidth]{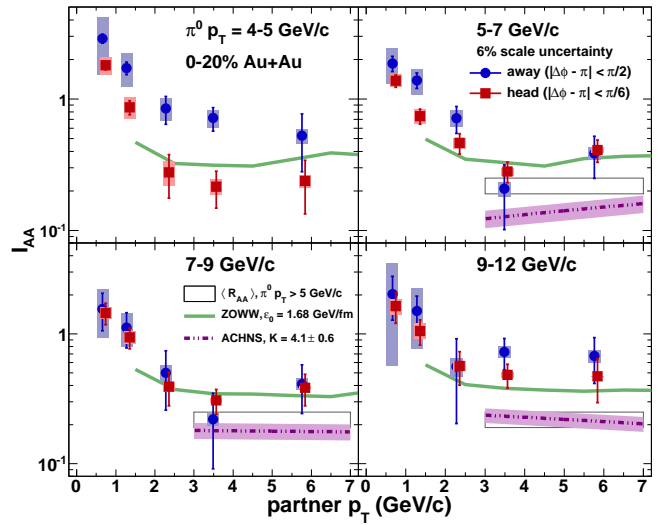}
  \caption{\label{fig:iaa} (Color online) Away-side $I_{\rm AA}$ for a
    narrow ``head'' $|\Delta\phi-\pi|<\pi/6$ selection (solid squares)
    and the entire away-side, $|\Delta\phi-\pi|<\pi/2$ (solid circles)
    vs. $h^\pm$ partner momentum for various \p trigger
    momenta. Calculations from two different predictions are shown for
    the head region in applicable \pt ranges. A point-to-point
    uncorrelated 6\% normalization uncertainty (mainly due to
    efficiency corrections) applies to all measurements. For
    comparison, \p \raa \cite{Adare:2008qa} bands are included where
    \ptt $>$ 5 GeV/$c$. }
\end{figure}

In addition to the shape modification measurement, the away-side
integrated yield is determined.  Away-side jet yield modification
in central collisions, shown in Fig.~\ref{fig:iaa}, is measured by
$I_{\rm AA}$ (the ratio of conditional jet pair yields integrated over a 
particular range in $\Delta \phi$ in Au$\mathord{+}$Au to $p\mathord{+}p$).
The $I_{\rm AA}$ uncertainties include uncorrelated errors
($\sigma_{\rm stat}$), point-to-point correlated errors from the
background subtraction ($\sigma_{\rm syst}$), and a normalization
uncertainty from the single particle efficiency determination.
 
Away-side $I_{\rm AA}$ values for \ptt $>$ 7 GeV/$c$ tend to fall 
with \pta for both the full away-side region ($|\Delta\phi-\pi| < 
\pi/2$) and for a narrower ``head'' selection ($|\Delta\phi-\pi| < 
\pi/6$) until \pta $\approx$ 2--3 GeV/c, above which they become roughly 
constant. The yield enhancement at \ptt $>$ 7 GeV/$c$ and \pta $<$ 
2 GeV/$c$ is modest and occurs without significant shape 
modification (Fig.~\ref{fig:sigma}). When \ptt is decreased, the 
away-side $I_{\rm AA}$ differs between the two angular selections 
as the shape becomes modified.

Average away-side $I_{\rm AA}$ values from weighted averages of the 
``head'' region data in Fig.~\ref{fig:iaa} for $\ptt(\pta) > 
5(2)$ GeV/$c$ are listed in Table~\ref{tab:iaa} for both central 
and midcentral collisions. The fits, which are not shown, cover 
the momentum range where shape modification is weak or nonexistent. 
\begin{table}[tb]
\caption{\label{tab:iaa}
  Average away-side $I^{\rm head}_{\rm AA}$ above 2 GeV/$c$ 
  for various \p trigger momenta in central and midcentral 
  collisions where $|\Delta\phi-\pi|<\pi/6$.  
  Note: a 6\% scale uncertainty applies to
  all $I_{\rm AA}$ values.
}
\begin{ruledtabular} \begin{tabular}{ccccccc}
\multicolumn{1}{c}{ } & \multicolumn{3}{c}{Cent 0--20\%} 
& \multicolumn{3}{c}{Cent 20--60\%} \\
$p^{t}_{T}$ & $I^{\rm head}_{\rm AA}$ &
$\pm\sigma_{\rm stat}$ & $\pm\sigma_{\rm syst}$ & 
$I^{\rm head}_{\rm AA}$ & $\pm\sigma_{\rm stat}$ 
& $\pm\sigma_{\rm syst}$ \\
\hline
5--7  & 0.35 & 0.04 & 0.03 & 0.55 & 0.02 & 0.04 \\
7--9  & 0.34 & 0.05 & 0.03 & 0.64 & 0.04 & 0.02 \\
9--12 & 0.50 & 0.08 & 0.02 & 0.73 & 0.06 & 0.02 \\
\end{tabular} \end{ruledtabular}
\end{table}
The away-side $I_{\rm AA}$ values for both centrality selections 
tend to rise as \ptt increases. Reference~\cite{starhighpt} 
measured a constant away-side $I_{\rm AA}$ for $z_T$ 
($=p^{a}_T/p^{t}_T$) above 0.4 for triggers at 8--16 GeV/$c$, but 
such a single point spanning a broad momentum range fails to 
provide information on the \ptt evolution of \iaa for comparison 
with the present results.

Figure~\ref{fig:iaa} also shows the \p \raa for \pt $>$ 5 
GeV/$c$~\cite{Adare:2008qa}.  The comparison reveals that \iaa is 
consistently higher than \raa. This feature probably results from 
a few competing effects. Selection of a high \pt trigger \p is 
expected to bias the hard scattering towards the medium surface. 
Thus, away-side partons have a long average path length through 
the medium and consequently lose more energy.  However, this does 
not require that \iaa be lower than \raa. The away-side 
conditional spectrum falls less steeply than the inclusive hadron 
spectrum and so the same spectral shift will more strongly reduce 
\raa.

Figure~\ref{fig:iaa} also shows \iaa calculations from the 
ACHNS~\cite{Armesto:2009zi} and ZOWW~\cite{Zhang:2009rn} models.  
Each calculation includes the combination of 
a parton energy loss formalism and a modeling of medium geometry.  
The ACHNS calculation, which employs a hydrodynamic evolution 
model of the medium and an energy loss prescription based on 
quenching parameters constrained by other 
data~\cite{Adare:2008qa,starhighpt}, predicts \iaa $\lesssim$ 
\raa.  The ZOWW calculation, which utilizes a simple spherical 
nuclear geometry and is similarly constrained by other 
data~\cite{Adare:2008qa,starhighpt}, predicts \iaa $>$ \raa in 
agreement with these data. It would be instructive to re-calculate 
these \iaa predictions with a common medium geometry (as was done 
for \raa in Reference~\cite{Bass:2008rv}) to disentangle the model 
differences. Additionally, a full assessment including all \raa 
and \iaa measurements, including direct photon trigger 
data~\cite{ppg090,stardirectph}, is warranted.

In summary, \p-\h correlations over a very broad range in trigger 
and partner \pt have been measured.  We observe an away-side 
modification for moderate \pt triggers (\ptt $<$ 7 GeV/$c$) and 
low \pt partners (\pta $<$ 3 GeV/$c$) as has been observed in 
unidentified dihadron correlations. However, this modification is reduced 
or absent for triggers above 7 GeV/$c$ for any partner \pt and challenges 
descriptions where more (initially) energetic partons lose more 
energy and should produce a larger medium response. At large 
momenta, i.e. triggers above 5 GeV/$c$ and partners above 2 
GeV/$c$, away-side modification factor \iaa is above the inclusive \p 
modification factor \raa(\pt$>$ 5 GeV/$c$).



We thank the staff of the Collider-Accelerator and 
Physics Departments at BNL for their vital contributions.  
We acknowledge support from 
the Office of Nuclear Physics in DOE Office of Science, NSF,
and a sponsored research grant from Renaissance Technologies (USA),
MEXT and JSPS (Japan), 
CNPq and FAPESP (Brazil), 
NSFC (China), 
MSMT (Czech Republic),
IN2P3/CNRS and CEA (France), 
BMBF, DAAD, and AvH (Germany), 
OTKA (Hungary), 
DAE and DST (India), 
ISF (Israel), 
NRF (Korea), 
MES, RAS, and FAAE (Russia),
VR and KAW (Sweden), 
U.S. CRDF for the FSU, 
US-Hungary Fulbright, 
and US-Israel BSF.


\end{document}